\documentclass[12pt]{article}
\usepackage{epsfig,graphicx}

\title{Magnetic focusing in atomic, nuclear and hadronic processes}
\author{  Yu.A.Simonov,\\ Institute of Theoretical and Experimental
Physics\\ 117118, Moscow, B.Cheremushkinskaya 25, Russia}
\date{}

\newcommand{\be}{\begin{equation}}
\newcommand{\ee}{\end{equation}}

\def\la{\mathrel{\mathpalette\fun <}}
\def\ga{\mathrel{\mathpalette\fun >}}
\def\fun#1#2{\lower3.6pt\vbox{\baselineskip0pt\lineskip.9pt
\ialign{$\mathsurround=0pt#1\hfil ##\hfil$\crcr#2\crcr\sim\crcr}}}

\newcommand{{\SD}}{\rm SD}

\newcommand{\vex}{\mbox{\boldmath${\rm x}$}}

\newcommand{\ver}{\mbox{\boldmath${\rm r}$}}
\newcommand{\vesig}{\mbox{\boldmath${\rm \sigma}$}}

\newcommand{\veP}{\mbox{\boldmath${\rm P}$}}
\newcommand{\vep}{\mbox{\boldmath${\rm p}$}}

\newcommand{\veQ}{\mbox{\boldmath${\rm Q}$}}

\newcommand{\veA}{\mbox{\boldmath${\rm A}$}}

\newcommand{\vek}{\mbox{\boldmath${\rm k}$}}

\newcommand{\veta}{\mbox{\boldmath${\rm \eta}$}}
\newcommand{\veB}{\mbox{\boldmath${\rm B}$}}

\newcommand{\veE}{\mbox{\boldmath${\rm E}$}}

\newcommand{\vepi}{\mbox{\boldmath${\rm \pi}$}}

\newcommand{{\Mc}}{\mathcal{M}}

\newcommand{\lan}{\langle}
\newcommand{\ran}{\rangle}

\begin{document}

\maketitle
\begin{abstract}
Processes with oppositely charged spinor particles in initial
and/or final states in homogeneous  magnetic field B are subject
to focusing effects in their relative motion, which yield the
amplifying factors in probabilities growing as $eB$. In addition
the increasing energy of some Landau levels influences the phase
space. As a result some processes in the proper spin states can be
enlarged as $\sim \frac{eB}{\kappa^2}$, where $\kappa^2$ is the
characteristic 2d phase space factor available without magnetic
field. Several examples, including neutron $\beta$ decay,
positronium decay and $e^+e^-$ pair production, are quantitatively
considered.

\end{abstract}

\section{Introduction}

The motion of charged particles in magnetic field (m.f.) is a standard topic of
textbooks \cite{1,2,3}, and the behavior of atomic and nuclear systems in the
framework of QED is extensively studied \cite{4,5}. Recently also the hadronic
systems in m.f. have attracted a lot of attention
\cite{7,8,9,10,11,12,13,14,15,16,17}. In particular, the role of m.f. in chiral
symmetry breaking (CSB) was stressed both analytically \cite{7} and on the
lattice \cite{8}, see \cite{9} for  a review and references, and the
corresponding phenomenon was coined   magnetic catalysis.

The dynamical origin of magnetic catalysis in QCD was discovered recently in
\cite{18} and shown  to be an example of a more general phenomenon  -- the
magnetic focusing, which assembles together particles of  opposite charges.

The study of relativistic QCD systems (quarks, gluons, hadrons) in m.f. has
made  it necessary to create and exploit the relativistic formalism, based on
the path integrals with interaction as in Wilson loops, which allows to write
down simple form Hamiltonians incorporating electromagnetic and strong
interactions \cite{10}.

This formalism was used  recently to calculate spectra of mesons in m.f.
\cite{11, 12, 13}, including the Nambu-Goldstone mesons \cite{14}, and meson
magnetic moments \cite{15}.

In the course of these studies it was found, that m.f. plays a very important
role in ``assembling'' opposite charges near one another, i.e. the ``focusing
effect'' to give giving contribution to the hyperfine(hf) splitting $\sim |\psi (0)|^2$
growing as $eB$ in hydrogen \cite{16}, as well as in relativistic $q\bar q$
systems \cite{17}, making it necessary to introduce the smearing effect to
consider hf as a perturbation. Moreover, it was found in \cite{18}, that the
characteristic growth of quark condensate $|\lan \bar q q\ran|$ with $eB$ is
again due to the fact, that it is proportional to $|\psi(0)|^2\sim eB$, i.e.
the focusing inside the $q\bar q$ pair.

It is clear, that the focusing mechanism is of a general  character and should
show up in all cases, where such a factor  $|\psi(0)|^2$ for the wave function
of relative coordinate of two oppositely charged particles appear. From the
general scattering theory \cite{19} it was  shown, that this factor (for
orbital  momentum zero) always appears whenever the reaction has two strongly
different ranges,  $r_{\rm ext} \gg r_{\rm int}$ \be  dw = |\psi^{(f)}_{\rm ext
(r {\rm int})} |^2 dw_{\rm int} |\psi_{\rm ext}^{(i)}(r_{\rm int})|^2
\label{1m}\ee  where superscripts $f, i$ refer to final, initial states. These
effects of initial state interaction (ISI) or final state interaction (FSI)
were carefully studied for the combination of Coulomb and nuclear forces
\cite{20}.

In the case of m.f. two differences appear:

\begin{enumerate}
    \item  m.f. induce 2d discrete spectrum in external motion, hence the sum
    over spectrum should enter in (1) instead of a simple  factor $|\psi_{\rm
    ext} (0)|^2$.
    \item  The masses of this discrete spectrum are  generally growing with $eB$ and
    strongly influence  the available  phase space, making in some  cases
     the process impossible.

\end{enumerate}

However for some lowest Landau levels (LLL),the energy is   \be
E_{n_\bot} = \sqrt{ m^2 + (2n_\bot + 1 -\sigma ) eB +
p^2_z},\label{2m}\ee with $n_\bot =0$, where  spin (magnetic
moment) contribution $\sigma =1$  exactly  cancels the radial
motion,yielding $E_0 = \sqrt{ m^2 + p^2_z}$.Thus one can  gain in
the resulting  energy and phase space and one obtains for the
creation of a pair the factor of growth \be \rho (eB) =
\frac{w(eB)}{w(0)} = \frac{eB}{\kappa^2}\label{3m}\ee where
$\kappa^2$ is the characteristic phase space, available for the
perpendicular relative motion of two charges without m.f.

Our subsequent discussion in section 2 will be using the relativistic
Hamiltonians in m.f. and the  resulting eigenfunctions and energies, obtained
in \cite{10,11,12,13,14,15} and applicable in QED and QCD, augmented by the
appropriate interaction terms. We shall proceed in section  3 with the simple
example of $e^+e^-$ production by $\gamma$ in some reaction, and then comparing
it with the pair creation in the  constant electric field.

In section 4 we turn to  the 3 body final state and consider the neutron
$\beta$ decay in m.f.

Other possible systems are discussed in section 5.  In section 6 we give  a
short summary and prospectives. The appendix contains a short derivation of Eq.
(\ref{1m}) using Jost solutions for external interaction.

\section{ Relativistic and nonrelativistic dynamics in strong magnetic field}

Our final goal is to demonstrate how m.f. changes the relative  motion of
opposite charged  particles and leads to the enhancement of the corresponding
wave function at small distances, thus yielding the amplification factor for
annihilation or production of such particles -- the phenomenon of magnetic
focusing.

To this end we are writing the relativistic Hamiltonian for a pair of particles
with charges $e_1 = - e_2 \equiv e$, which was already derived and exploited in
\cite{11,12,13} and rederived in the framework of the new path integral
representation in \cite{10}. For particles with masses $m_1, m_2$ in m.f.
$\veB$ along $z$ axis the Hamiltonian has the form (after the proper
pseudomomentum factorization \cite{10,11,12}).

\be H= \frac{\veP^2}{2(\omega_1 + \omega_2)} + \frac{\vepi^2}{2\tilde\omega} +
\frac{1}{2\tilde \omega} \frac{e^2}{4} (\veB\times \veta)^2 + \sum_{i=1,2}
\frac{m^2_i+ \omega_i^2- e_i \vesig_i \veB}{2\omega_i} + \hat V.\label{1}\ee

Here $\veta = \ver_1 - \ver_2, ~\vepi = \frac{\partial}{i\partial\veta} =
\frac{\omega_1 \vek_1 - \omega_2 \vek_2}{\omega_1 +\omega_2},~~ \tilde \omega=
\frac{\omega_1 \omega_2}{\omega_1 + \omega_2}$, and $\hat V$ is the sum of all
interaction terms, including photon or gluon exchange $V_{\rm Coulomb} \equiv
V_c (\veta)$, confining interaction $V_{\rm conf}$ for quarks, and
spin-dependent and self-energy corrections, for details see \cite{17}.

The eigenfunction $\Psi(\omega_1, \omega_2 )$ and eigenvalues $M_j\equiv
M_{n_\bot, n_z} (\omega_1, \omega_2)$ depend on $\omega_1, \omega_2$ and the
actual energy eigenvalue $M^{(0)}_j$ is obtained from $M_j (\omega_1,
\omega_2)$ by the stationary value procedure, \be M^{(0)}_j = M_j
(\omega_1^{(0)}, \omega_2^{(2)}), ~~ \left.\frac{\partial M_j (\omega_1,
\omega_2)}{\partial \omega_i}\right|_{\omega_i = \omega_i^{(0)}} =
0.\label{2}\ee This scheme is discussed in detail in \cite{10}.

For nonrelativistic approximation the dominant terms are $\frac{m^2_i +
\omega^2_i}{2\omega_i}$,which automatically give $\omega_i^{(0)} = m_i$.

For strong m.f., when one can neglect $\hat V$ in (\ref{1}), i.e. for $eB\gg
\sigma$ in hadron systems and $eB\gg (m_e\alpha)^2$ in atomic systems ,one
immediately obtains the c.m. values of $M_j^{(0)}$ \be M_j^{(0)}  = \sqrt{
m^2_1 + \pi^2_z + eB (2n_\bot + 1 - \sigma_{1z})} + \sqrt{m^2_2 + \pi^2_z + eB
(2n_\bot + 1+\sigma_{2z})}\label{3}\ee and the eigenfunction for $n_\bot =0$ is
(neglecting Coulomb interaction) \be \Psi (z, \veta_\bot) =
\frac{e^{i\pi_zz}}{\sqrt{L}} \varphi_{n_\bot} (\veta_\bot), ~~ \varphi_{n_\bot}
(\veta_\bot)= \frac{e^{-\frac{\eta^2_\bot}{2r^2_\bot}}}{\sqrt{\pi} r_\bot}, ~~
r_\bot = \sqrt{\frac{2}{eB}}, ~~ \varphi^2_0 (0) = \frac{eB}{2\pi}.\label{4}\ee

Note, that $\varphi_0^2 (0) $ grows linearly with $eB$, the property, which is
the basic for the magnetic focusing phenomenon. As will be seen, another
property is important: the energy eigenvalue $M^{(0)}_{n_\bot=0}
(\sigma_{1z}=+1, \sigma_{2z}=-1)$ does not grow with (and does not depend on)
$eB$. Thus the Lowest Landau Level (LLL) with $n_\bot =0$  and $\sigma_{1z} =-
\sigma_{2z} =1$ ( we shall call it the ``zero level'') ensures the enhancement
of probability of production or annihilation of the pair without increase of
the spectrum, which would otherwise stop the process.

\section{The  electron-positron pair production in m.f.}

Consider the process $A+B\to C + (e^+e^-)$, where $e^+e^-$ is
produced by a virtual photon. The amplitude for the process
without m.f. can be written as \be \mathcal{M} =C_\mu
\frac{1}{Q^2} (\bar \psi \gamma_\mu \psi), ~~ \psi=
\frac{u}{\sqrt{V_3}} e^{ik_+ x},~~ \bar \psi = \frac{\bar u
e^{-ik_-x}}{\sqrt{V_3}} \label{5}\ee and one should have in  mind
that  both $e^+$ and $e^-$ are created at one point $x$. The
probability can be written as
$$dw =\left| C_\mu \frac{(\bar u \gamma_\mu u)}{Q^2}\right|^2
\frac{d^3k_+d^3k_-}{(2\pi)^6} \delta^{(4)} (Q-k^+-k^-)(2\pi)^4=$$ \be =\left|
C_\mu \frac{(\bar u \gamma_\mu u)}{Q^2}\right|^2\frac{d^3k_+}{(2\pi)^2} \delta
(Q_0- 2\sqrt{(\vek^+)^2+m^2_e}),\label{6}\ee where we have assumed, that
$\veQ=0$.

We now turn to the case of nonzero m.f.Then the energy of $e^+e^-$ in m.f.
$\veB$ can be written, using (\ref{3}) as

\be E_{n_\bot}(\pi_z,B)  = \sqrt{ m^2_e + \pi^2_z + eB (2n_\bot + 1 -
\sigma_{+z})} + \sqrt{m^2_e + \pi^2_z + eB (2n_\bot +
1+\sigma_{-z})}\label{7}\ee where $\pi_z= \frac{k_{+z}-k_{-z}}{2} = k_{+z}$,
and $\sigma_{+z}$ and $\sigma_{-z}$ are doubled spin projection of $e^+$ and
$e^-$ respectively, and we take into account, that $e_+ = - e_-\equiv e$.

For the probability one can write \be dw =\left| C_\mu \frac{(\bar u \gamma_\mu
u)}{Q^2}\right|^2 \frac{d\pi_z }{2\pi} \sum_{n_\bot =0} \delta  (Q_0-
E_{n_\bot} (\pi_z))\varphi^2_{n_\bot} (0),\label{8}\ee  and again $\varphi^2_0
(0) = \frac{eB}{2\pi}$, as in (\ref{4}).

It is important, that for  the lowest energy state at $eB\to \infty$ both terms
$(1-\sigma_{1z})$ and $(1+\sigma_{2z})$ should vanish, in which case the term
with $n_\bot =0$ always survives in (\ref{8}), since $E_0$ does not contain
$eB$. Looking at the structure $\bar u \gamma_\mu u$ and taking into account,
that $u_+ = C\bar u, ~~ C= \gamma_2 \gamma_4$, one comes to the conclusion,
that for $\gamma_\mu = \gamma_3$ the cancellation mentioned above  is possible,
while for $\gamma_\mu =\gamma_1, \gamma_2, \gamma_4$ both $\sigma_{+z} $ and
$\sigma_{-z}$ have the same sign. Therefore for $eB\to \infty$ only the term
with $\gamma_3$ survives and one obtains the behavior.\be dw
\cong\frac{eB}{2\pi} \left| C_3 \frac{(\bar u \gamma_3 u)}{Q^2}\right|^2
\frac{d\pi_z }{2\pi} \delta  (Q_0- E_{0} (\pi_z)), \label{9}\ee which
demonstrates the linear growth of the pair production with $eB$.

In the standard calculation of the $e^+e^-$ pair production in m.f. (see e.g.
\S 91 of \cite{3}, and original papers \cite{19*} the probability was
calculated using crossing relations with the process $e^- \to e^- +\gamma$ in
m.f. and therefore the magnetic focusing effect was not taken into account.

It is interesting to apply our results to the process of $e^+e^-$ pair creation
in the constant electric field (in Minkovski space-time).

In the proper-time formalism \cite{21}, see also chapter 4 of \cite{4}, the
pair probability is \be w(x) = Re ~tr \int^\infty_0 \frac{ds}{s} e^{-is
(m^2-i\varepsilon) } \lan  x | (e^{is \hat D^2}- e^{is\hat
\partial^2})|x\ran\label{11}\ee
and $D_\mu= \partial_\mu - ie A_\mu(x), ~~ A_3 (x) = - Et, t=x_0$, $\veA_\bot =
\frac12 (\veB \times \vex), \veB$ is along $z$ axis.

According to our discussion above, only in this situation, when
$\veE\parallel\veB$, the factor $\gamma_3$ in $\hat A$ generates zero levels
and the magnetic focusing produces the factor $\frac{eB}{2\pi}$ for each pair
without increasing effective mass of the $e^+e^-$ pair.  The explicit form of
the pair-production rate  per volume per time was found in \cite{20A} for
parallel $\veE$ and $\veB$ (remember $\gamma_3$ reasoning above) \be w=
\frac{(eE) (eB)}{(2\pi)^2} \coth \left( \frac{\pi B}{E}\right) \exp
\left(-\frac{\pi m^2}{eE}\right),\label{13*}\ee where for $\pi B\gg E$ one can
see a linear growth of $w$ with increasing $eB$.  See more on pair creation in
\cite{4}.

Now comparing (\ref{6}) and (\ref{8}) one finds the correspondence \be
\frac{d^2\pi_\bot}{(2\pi)^2} \to \sum_{n_\bot =0}^{n_\bot(\max)}
\varphi^2_{n\bot} (0)\label{12}\ee.

At this point one can compare our results with the statistical
weight argument suggested in \cite{1,2}, where the substitution
should be performed with inclusion of magnetic field \be \int
\frac{d^3p}{(2\pi)^3} f (E) \to \int \sum^\infty_{n_\bot =0}
\frac{dp_z}{2\pi} f (E(B))\frac{eB}{2\pi} \label{11*}\ee and
the sum over spin projections is assumed. One can see a close
correspondence between (\ref{8}) and (\ref{11*}), however
(\ref{11*}) has to be attributed to any final particle,  and only
quasiclassical arguments are used for (\ref{11*}) in \cite{1,2}.

It is interesting how the form (\ref{8}) goes over into (\ref{6}) when $eB\to
0$. To this end we write, generalizing $\varphi^2_{n\bot} (0)$  to
$|\varphi_{n\bot} (r_{\rm int})|^2$,

$$ \varphi_{n_\bot} (r_{\rm int} ) = \int \tilde \varphi_{n_\bot} (\vepi)
\frac{d^2\vepi}{(2\pi)^2}\exp (i\vepi \ver_{\rm int})$$ and \be \sum_{n_\bot
=0}^{n_\bot(\max)} |\varphi_{n_\bot} (r_{\rm int})|^2= \sum_{n_\bot
=0}^{n_\bot(\max)} \int\int\frac{d^2\vepi}{(2\pi)^2}\frac{d^2\vepi'}{(2\pi)^2}
\tilde \varphi_{n_\bot} (\vepi )\tilde \varphi_{n_\bot}^* (\vepi'
)e^{i(\vepi-\vepi') \ver_{\rm int}}.\label{13}\ee

For large $n_\bot (\max)\gg 1$ one can use the property of completeness of the
set $\{\tilde \varphi_{n_\bot} (\vepi )\}$, which yields \be
\sum^\infty_{n_\bot =0} \tilde \varphi_{n_\bot} (\vepi )\tilde
\varphi_{n_\bot}^* (\vepi ')= (2\pi)^2 \delta^{(2)}
(\vepi-\vepi'),\label{14}\ee and \be \sum_{n_\bot =0}^{n_\bot(\max)}
|\varphi_{n_\bot} (r_{\rm
int})|^2\cong\int^{\vepi(\max)}_0\frac{d^2\vepi}{(2\pi)^2}.\label{15}\ee

In this way we are proving the correspondence (\ref{12}), (\ref{11}) for any
$r_{\rm int}$; note, however, that we have accounted only for the final state
interaction (FSI) in m.f. and have shown, that it can be equivalently written
in the form of a modified  statistical weight. However,  in the sum over
$n_\bot$, the factor $|\varphi_{n_\bot}(r_{\rm int})|^2$ does not reduce to
$\frac{eB}{2\pi}$, when $eB> \frac{1}{r_{\rm int}^2}$.

\section{Formalism for 3 body final states}

In this section we consider an example of the neutron $\beta$ - decay, $ n\to p
+ e^-+\bar \nu_e$, in the presence of the homogeneous magnetic field $\veB$
along $z$ axis

Writing the matrix element as \be \mathcal{M}_{if} = G\cos \theta (\bar \psi_p
(x)O^{(1)}_\mu\psi_n (x)) (\bar \psi_e(x) O^{(2)}_\mu \psi_\nu), \label{16}\ee
where $O_\mu^{(i)} = \gamma_\mu (1-\alpha_i \gamma_5)$, and $G\cos \theta\equiv
\bar G  = \frac{1.0 \cdot 10^{-5}}{m^2_p}, \alpha_1=1,262, \alpha_2=0$ and  for
$B=0$ $\psi_k(x) = \frac{ue^{i\vep_k\vex}}{\sqrt{V_3}}$, one obtains for the
decay probability $$ dw = \bar G^2 |(\bar u 0_i u) (\bar u0_iu)|^2\left(
\prod_{k=2,3,4} d^3k\right) \frac{\delta^{(4)} (p_1-p_2-p_3-p_4)}{(2\pi)^5}$$
\be = \bar G^2|(\bar u 0_i u) (\bar u0_iu)|^2 \frac{(M_n -
\varepsilon_2-\varepsilon_3)\varepsilon_2 d\varepsilon_2 \varepsilon_3 d
\varepsilon_3}{(4\pi)^3}\label{17}\ee and $\varepsilon_i = \sqrt{\vep^2_i+
m^2_i}, ~~ i=2,3$ refers to proton and electron respectively.

The integration in (\ref{17}) proceeds inside the phase space of $ep$ system.

We now want to apply the m.f. to our system and to this end we realize that it
will act only on the relative coordinate $\veta_\bot $ of ($ep)$  system,
perpendicular to the  $\veB$ direction, and we must separate out the
$\veta_\bot$ dependence  both in the $(ep)$ wave function and in  the phase
space integration one can consider (\ref{16}) as a matrix element of the
operator $\hat T$ between initial and final states, see appendix,
$\mathcal{M}_{if} = \lan \Psi^+_f| \hat T|\Psi_i\ran$, and both $\Psi_f$ and
$\Psi_i$ enter at one space-time point in the limit of large $W$ mass.

Therefore $\Psi_f(x) = \psi_p (x) \psi_e(x) \to \Psi_{ep}(x)$ and the latter
w.f. is  defined by  the m.f. Hamiltonian (\ref{1}).

According to \cite{5}, the Hamiltonian for the neutral ($ep)$ system can be
written as in (\ref{1}) with $\omega_1 = \omega_p\approx m_p, ~~ \omega_2 =
\omega_e \ll m_p$.

Noting, that $\omega_1 \equiv \omega_p \approx m_p \gg\omega_2 \equiv
\omega_e$, one obtains  $$E^{(0)}_{ep} \equiv E_{ep} (\omega_p^{(0)},
\omega_e^{(0)} ) \cong \frac{\veP^2}{2m_p} +$$\be+ \sqrt{m^2_p + \pi^2_z + eB
(2n_\bot +1-\sigma_{pz})}+\sqrt{m^2_e + \pi^2_z + eB (2n_\bot
+1-\sigma_{ez})}.\label{18}\ee

Now the  phase space can be  rewritten as follows \be\frac{d^3p_p
d^3p_e}{(2\pi)^6} =\frac{d^3P d^3\pi}{(2\pi)^6} =\frac{d^3P d\pi_z}{(2\pi)^4}
\frac{d^2\pi_\bot}{(2\pi)^2}\label{19}\ee and $$ d\Phi\equiv \frac{d^3p_\nu
d^3p_p d^3p_e}{(2\pi)^5} \delta^{(4)} ( p_n-p_p-p_e-p_\nu)=$$ \be
=\frac{d^3Pd\pi_z d^2\pi_\bot}{(2\pi)^5} \delta (m_n - |\veP| - E_{ep}^{(0)}) =
\frac{P^2d\pi_z}{2\pi^2 \left(1+ \frac{P}{m_p}\right)}
\frac{d^2\pi_\bot}{(2\pi)^2}\label{20}\ee writing $E_{ep}^{(0)} \cong m_p +
\varepsilon_e, \varepsilon_e \equiv \varepsilon$,one can express (\ref{17}) in
the form:
 \be d\Phi= \frac{(\Delta m- \varepsilon)^2}{4\pi^3}
d\pi_z \pi_\bot d\pi_\bot = (\Delta m-\varepsilon)^2 \frac{\sqrt{\varepsilon^2
- m^2_e}}{2\pi^3} \varepsilon d\varepsilon; ~~ \Phi= \int d\Phi \cong
\frac{1.63}{2\pi^2} m^5_e\label{21}\ee$\Delta m = 1.29$ MeV, and finally \be
w_n = G^2 (1+3\alpha^2) \Phi; ~~ \alpha\equiv g_A/g_V \approx
1.262\label{22}\ee In the presence of m.f. the values of $\pi_\bot$ are
quantized in the Hamiltonian (\ref{1}), so that one should replace as in
(\ref{12}), so  that finally the decay probability becomes \be dw = g^2|(\bar u
0_i u) (\bar u0_iu)|^2\frac{P^2d\pi_z}{4\pi^2 \left(1+
\frac{P}{m_p}\right)}\sum^{n_\bot(\max)}_{n_\bot=0}
\varphi^2_{n_\bot}(0)\label{23}\ee and  $n_\bot(\max)$ is defined by the
condition $m_n = P+E^{(0)}_{ep} (n_\bot)$.

As was shown in (\ref{15}) for $eB\to 0$ one has the answer (\ref{17}).

In  the opposite limit, when $eB$ is large, $eB\ga m^2_e$ and $n_\bot (\max)
=0$, one can read in (\ref{18})  that $E_{ep}^{(0)}$ is \be E_{ep}^{(0)} (eB
\to \infty) \cong m_p + \sqrt{m^2_e+ \pi^2_z}, ~~ \sigma_{ez} =-1\label{24}\ee
and \be \Delta m \equiv m_n - m_p = P + \sqrt{m^2_e+ \pi^2_z}\label{25}\ee
which defines allowable phase space  for $(P, \pi_z)$. In  this case for
$n_\bot =0$ one has the amplifying factor as in (\ref{4}) $ \varphi_0^2 (0) =
\frac{|eB|}{2\pi}, $ which can be much larger, than $\int^{\pi_\bot(\max)}
\frac{d^2\pi_\bot}{(2\pi)^2} = \frac{\pi^2_\bot(\max)}{4\pi} \leq\frac{(\Delta
m)^2 - m^2_e}{4\pi}$ for $eB \gg (\Delta m)^2-m^2_e$, see \cite{20*,20**}.

Inserting the values of $\Delta m =  2.53 m_e$, one obtains the amplifying
factor for $eB\gg m^2_e$, when only one term $n_\bot =0$ should be kept in
(\ref{22}),  $\frac{w_n(eB)}{w_n(0)} = \frac{eB}{\kappa^2}$,and $\kappa^2\approx
m^2_e$. Exact calculation in \cite{20*} yields $\frac{w_n(eB)}{w_n(0)} = 0.77
\frac{eB}{m^2_e}$.

Note, however, that for small $\pi_\bot (\max) $  and hence small $\kappa^2$,
this ratio can be arbitrarily large.

For more discussion of this subject and additional references see \cite{24*}.

\section{Other interactions and other systems}

One immediate application of the magnetic focusing is the hyperfine interaction
in all systems. In hydrogene this effect was studied in \cite{16} and it was
shown, that in the standard form of the hyperfine shift

\be \Delta E_{hf} =\frac{32 \pi}{3} g_p \mu_B \mu_n | \Psi (0) |^2\label{26}\ee
the wave function of the ground state can be chosen as \be \Psi_0 (\eta_\bot,
z) = N\exp \left( - \frac{\eta_\bot^2 a^2}{2} -
\frac{z^2b^2}{2}\right)\label{27}\ee with $|\Psi(0)|^2 = \frac{a^2
b}{\pi^{3/2}}$, and $a,b$ are fitting parameters; for the free election
$a=\frac{eB}{2m_e}$, and for large $eB> (m_e \alpha)^2, a^2b$ grows with $eB$
almost linearly, which gives the possibility to study this effect
experimentally.

The same effect was found in relativistic hadronic systems in \cite{17}, where
again the hf shift has the same form as in (\ref{21}), namely \be \lan
V_{hf}\ran = \frac{8\pi\alpha_s (\vesig_1\vesig_2)}{9 \omega_1\omega_2} |
\Psi_{q\bar q}(0)|^2\label{28}\ee and one can show, that (\ref{28})  when
$\omega_1, \omega_2$ are decreasing and $\vesig_1 \vesig_2=-3$ leads to the
absurd result of the negative PS masses at large $eB$, contradicting the
stability theorem, proved in \cite{17}, which tells, that energy eigenvalues in
magnetic and vacuum Euclidean fields cannot be negative and hence  $V_{hf}$
cannot be treated perturbatively, when $\omega_i \to 0$ and $\lan V_{hf}\ran
\to \infty$, so that  the relativistic smearing  must be  introduced, replacing
$\delta^{(3)} (\veta) $ in $V_{hf}$ by $\left(
\frac{1}{\mu\sqrt{\pi}}\right)^3\exp (-\mu^2\eta^2), $ $ \mu\approx (1\div 2)$
GeV.

In any case, the behavior (\ref{28}) signifies the strong increase of the
hyperfine splitting for hadrons in magnetic field, which is stabilized by the
smearing \cite{17}.

The calculations of hadron masses, subject to strong hf interaction have been
done in \cite{11,12,13,14}. In the case of $\pi^0, \rho^0$ mesons the  large
$eB$ asymptotics corresponds to $\lan \sigma_1 \sigma_2| = \lan +-|$ state
 for $\pi^0$ and $\lan -+|$ state for $\rho^0$ and the strong $hf$ interaction
 splits the masses, creating a deep minimum for the $\pi^0$ mass. However, this
 calculation in \cite{11} refers to the purely $q\bar q$ components of $\pi^0$,
 whereas the chiral dynamics in m.f. needs a special treatment, which was
 performed in \cite{13} and again showing a decreasing with $eB$ $\pi^0$ mass, with a minimum.

The calculation  in \cite{14},  was done basing on the unified theory of chiral
dynamics with $q\bar q$ degrees of freedom, developed earlier in \cite{22}. In
its turn in \cite{18} it was shown, that the chiral condensate grows  with
m.f., first quadratically for $eB < \sigma$ and then for $eB\gg \sigma$
linearly with $eB$. This behavior found in \cite{18} agrees quantitatively with
lattice data of \cite{8}, which contradict  CPTh. In this way one can conclude,
that the standard chiral theory can be applied only for $eB\la m^2_\pi$, as was
noticed before  in \cite{23}.

The subsequent analysis of Nambu-Goldstone (NG) mesons in m.f. was done in
\cite{14}, where it was shown, that GMOR relations are valid for neutral NG
mesons, but are violated for the charged ones, and  the $\pi^+$ mass was
calculated  in agreement with lattice data from \cite{24}. One should stress,
that the phenomenon of ``magnetic catalysis'', discussed in \cite{7,9} and
implying  the growth of chiral condensate $\lan \bar q q\ran$ in m.f. actually
occurs due to magnetic  focusing,  as was shown in \cite{18}, since $\lan \bar
q q\ran \sim \psi^2 (0) \sim eB$.

We now turn to other processes, where the opposite charge particles appear in
the initial state. One example is the $\mu^- + p \to n + \nu$. In the case,
when the process occurs from the ground state of $(\mu^-p)$ atom, one can use
the form (\ref{22}) with $r_\bot, r_z$ obtained in \cite{16}, which are reduced
by the ratio $\frac{m_e}{\tilde m_\mu}, \hat m_\mu= \frac{m_\mu m_p}{m_\mu=
m_p}$.

Since the  probability $w_\mu \equiv w(\mu^- + p \to n+\nu)$ is proportional to
$|\Psi_{\mu p} (0) |^2,$ one can obtain the amplification coefficient \be
\rho(B)\equiv\frac{w_\mu (B)}{w_\mu(0)} = \frac{r^2_\bot (0) r_z(0)}{r^2_\bot
(B) r_z (B)}, ~~ \rho(0)=1.\label{29}\ee

Using the data from the Fig.2 of \cite{16}, one easily obtains for $H=
\frac{eB}{\tilde m^2_\mu \alpha^2},~~ 0\leq H \leq 2$

$$ \rho (H=1) = \frac{1.33^3}{1^2 1.15} \cong 1.05; ~~\rho(H=2) =3.35.$$

One can see, that in this case one needs much larger fields $\left( \left(
\frac{\tilde m_\mu}{m_e}\right)^2~ {\rm times~ larger}\right)$ to produce the
same kind of effect, as in the electron case.

Let us turn now to the case of the positronium annihilation in m.f.  Since the
latter does not  conserve the spin, but only total spin projection $S_z$, it is
convenient to discuss separately the case  a) of $S_z=0$ parapositronium and
orthopositronium, and b) $S_z\pm 1$ orthopositronium. In the first case, as in
the $q\bar q$ case, both states are mixture of $\alpha =(\sigma_z = +1,
\sigma_z =-1) \equiv <+- |$ and $\beta = \lan -+|$, and parapositronium in
strong m.f. tends to be a pure $\alpha$ state, whereas orthopositronium a pure
$\beta$ state (see \cite{11,13} for a similar discussion in the $q\bar q$
case). For the case b) $S_z = \pm 1,$  in orthopositronium the corresponding
ground state masses in strong m.f. grows as $\sim \sqrt{eB}$ with additional
amplification due to the strong hf contribution. As a result the decay phase
space increases with $eB$ and the decay probability in the case b) is growing
both due to phase space and as in (\ref{29}), where in $H$ $\tilde m_\mu
\to\frac12 m_{e}.$ The same happens in the $\beta$ state of orthopositronium,
since the total energy in the $\beta$ state also grows as $\sim \sqrt{eB}$.

However, in the $\alpha$ state of parapositronium the mass is
slightly decreasing, while the $|\psi(0)|^2$ factor is growing as
$eB$, and we expect the same situation, as in the example of
$(e^+e^-)$ pair creation in m.f., discussed in section 2 . At this
point it is necessary to stress also the difference between the
two examples: for positronium annihilation the factor $\rho(B)$ is
proportional to $| \psi(r_{an})|^2,$ , while the $(e^+e^-)$ pair
creation occurs at one point and brings about factor
$|\psi(0)|^2$. The resulting difference is expected to be of the
order of unity for $eB r^2_{an}\la 1$.

However for large $eB> (m_e\alpha)^2$ one expects the behavior to be $w_{+-}
(B) \approx \int|\psi_{+-} (r_{\rm an}) |^2 w^{(0)} d\tau,$  where $w^{(0)} d
\tau$ is the phase space integration (depending on $B$) with the two-photon
annihilation amplitude  $A^{(2)}, w^{(0)} = |A^{(2)}|^2$. Note, That both
$<+-|$ and $ <-+|$ states can  be superpositions of $S=1$ and $S=0$ states. As
a result we show the distributions $w_{+-} (B)$ and $w_{-+}(B)$ as functions of
$eB$ and notice, that the linear growth for $eB \gg(m_e\alpha)^2$ is saturated
for larger m.f., as was discussed previously.

In \cite{26} the authors have used the gaussian form of positronium wave
function and found the almost linear growth of two-photon annihilation with
$eB$ for $eB\la 10^{13}$ Gauss: $w(H=1) = 3.35\cdot  10^{12} s^{-1}, $  $
w(H=3) = 12.3\cdot  10^{12} s^{-1}, $ $ w(H=1) = 3.35\cdot  10^{12} s^{-1}, $
$w (H+10) = 5.09 \cdot 10^{13} s^{-1}, $ $w(H=44) = 11.8\cdot  10^{13} s^{-1},
$ where $H=\frac{B}{10^{12}} $ Gauss. One can see, that for $B=10^{13}$ Gauss
the growth of $w$ is weakening.

\section{ Summary and discussion}

We have demonstrated, that the acting of m.f. on the system of two opposite
charges in the initial or final state produces an amplifying factor, which can
 grow linearly with $eB$ for $eB$ in the range $(eB)_{\min} \la eB \la
 (eB)_{\max}$.

 In particular, for the $(e,-e)$  continuum production the relative probability
 $w(eB)/ w(0) \simeq |\psi_{cont}(r_{(int)}|^2/\kappa^2$ can grow as
 $\frac{eB}{\kappa^2}$, where $\kappa^2$ is the  effective phase space for the
 perpendicular motion in the $B=0$ case. In this case $(eB)_{\min}$ is also of
 the order of $\kappa^2$, while $(eB)_{\max}$ is defined by the
 $(eB)_{\max}\sim 1/r^2_{int}$, where $r_{int}$ refers to the range of the
 internal production process (e.g. $r_{int} \sim 1/m_W$ for neutron
 $\beta$-decay).

 We have illustrated this behavior by the processes of the $e^+e^-$ pair
 creation and the neutron $\beta $- decay; in the last case this amplification
 was known and calculated by many authors, see e.g.\cite{20*}, \cite{20**}, \cite{24*} using
 the quasiclassical calculation of phase  space (QPS) in m.f. as in \cite{1,2}.

 We have shown, that the FSI method of our paper gives the same result as QPS
 for $r_{int} =0$ or for $eB\ll 1/r^2_{int}$; however for $eB \sim 1/
 r^2_{int}$ the QPS  prediction overestimates the probability.

 For the $(e,-e)$ bound state production the amplification factor is
 proportional to $|\psi_{b.s.} (r_{int})|^2$, and for $r_{int} =0$ it grows
 linearly with $eB$ only for $eB\gg r^2_{b.s.}$, where $r_{b.s.}$ is the bound
 state radius at zero m.f., e.g. for positronium $r_{b.s.} = \frac{2}{m_e
 \alpha}$.

 The same  arguments can be applied in principle to the effect of constant m.f.
 in the initial state, as it is illustrated above in the paper by the example
 of the two-photon positron annihilation, see e.g. \cite{26}. However in this
 case one should distinguish different physical situations; in  astrophysics
 these processes are part of a  set of reactions in m.f. at finite temperature
 and chemical potential, while in  experiment one should take into account
 boundary conditions of the experimental device.

 Incidentally, one should stress, that all above derivations disregarded the
 Coulomb potential role in ISI and FSI.

 As it is known, the latter gives the factor for $(e, -e)$ system
 $|\psi_{coul}(0)|^2 = \frac{2\pi \eta}{\exp (-2\pi\eta) +1}, ~\eta
 =\frac{e^2}{v}$, which may bring an additional amplification of the magnetic
 focusing effect, discussed above. As it is, strictly speaking we can discuss
 the Coulomb amplification for $2\pi \eta\gg 1$ and $eB \ll \kappa^2$. however
 the explicit amplification factor in case, when both Coulomb interaction and
 m.f. are acting, is still not available and should be derived, using solutions
 when both interactions are present.

 Finally, probably the most important conclusion of our paper is, that for $(e,
 -e)$ systems in continuum there occurs an amplification factor
 $\frac{eB}{\kappa^2},$ which does not depend  on the masses of the charges and
 their sizes $R_i$ (provided $eB< 1/R^2$) and therefore can be important for
 stimulating different reactions in atomic, molecular, nuclear or hadronic
 reactions.
At this point one should compare our results for the process of
creation of the neutral system of two oppositely charged particles
with a similar process,where single charged particles appear.In
the latter case the particle is trapped in its motion across m.f.
and cannot reach detector at the distance R cm,when m.f. is higher
than approximately $1/R^2 \times 10^{-9}$ Gauss.This is in contrast
with our neutral system,which has in continuum spectrum the radius
${eB}^{-1/2}$ and is freely moving across and along m.f.,thus
yielding the finite cross section.
The author is grateful for useful discussions and suggestions to M.A.Andreichikov,B.M.Karnakov,
B.O.Kerbikov,V.D.Orlovsky and M.I.Vysotsky.

\vspace{2cm}
 \setcounter{equation}{0}
\renewcommand{\theequation}{A \arabic{equation}}

\hfill {\it  Appendix  }

\centerline{\bf \large Initial and final interaction factors in process
probability}

 \vspace{1cm}

\setcounter{equation}{0} \def\theequation{A \arabic{equation}}

One can use the formalism of {} to define the matrix element of a transition
from the state $\beta$ to $\alpha$, generated by the interaction $V$ in the
total hamiltonian \be H=K+U+V\label{A1}\ee where $K$ is the kinetic term and
$U$ and $V$ have different ranges $r_U, r_V$ correspondingly.  To the first
order in $V$ one can write:

\be f_{\beta\alpha} = (\varphi_\beta^- V \varphi^+_\alpha) +
O(V^2)\label{A2}\ee where $\varphi^-_\beta, \varphi^+_\alpha$ are ingoing and
outgoing solutions of the operator $K+U$, \be \varphi^-_\beta = \chi_\beta +
\frac{1}{E-K-i\varepsilon} U\varphi^-_\beta\label{A3}\ee and for
$\varphi^+_\alpha$ one should replace $\beta\to \alpha, \varepsilon \to -
\varepsilon$.

Introducing for the orbital momentum $l=0$ the internal amplitude due to $V$
only as $f_{in} (E)$, one can write (see \cite{19} for a detailed derivation)
\be f_{\alpha\alpha} = f_{in} (E) (\eta_{Jost})^{-2}\label{A4}\ee where
$\eta_{Jost} = \eta (0)$, and $\eta(r)$ is the Jost solution of the external
problem, satisfying the condition \be\lim_{r\to \infty} \exp (-ikr) \eta (r)=1,
\label{A5}\ee while for the Coulomb type interaction one should  replace
$-ikr\to -ikr+\frac{i\alpha}{v} \ln 2kr$.

 $\eta_{Jost}$ can be expressed via the regular solution
$\chi(r)$ of the external potential with the asymptotics \be \chi_{\rm ex}(r)
\sim \frac{\sin (kr+\delta_{\rm ext}) e^{i\delta_{\rm ext}}}{kr}, ~~ r\to
\infty,\label{A5a}\ee and  the connection is \be \lim_{r\to 0}\chi(r) =
\eta^{-1}_{Jost}.\label{A5aa}\ee and as a result one has (see \cite{19}), \be
f_{\alpha\alpha} (E) = \frac{f_{in}(E) (\chi_{\rm ex } (0))^2}{1+ f_{in} (E)
g(E, 0)}, \label{a5aaa}\ee where $g(e, r_{\rm in})$ is expressed via Green's
functions \be g(E,r) = G_0 (r, r_{\rm in} ) - G_{\rm ex} (r, r_{\rm
in})\label{A5aaaa}\ee

In particular for the Coulomb external problem one finds \cite{20}  \be
\eta_{Jost} = \exp \left( \frac{\pi \gamma}{2}\right)/ \Gamma(1+i\gamma) =
|\eta_{Jost}| \exp (-i\sigma_0),\label{A6}\ee

\be \exp (2i\sigma_0) =
\frac{\Gamma(1+i\gamma)}{\Gamma(1-i\gamma)},\label{A7}\ee

\be |\eta_{Jost}|^{-2} \equiv C^2_0 = \frac{2\pi\gamma}{\exp (2\pi\gamma) -1},
\label{A8}\ee where $\gamma = \frac{ze^2\mu}{k}$, and $\gamma=-|\gamma|$ for
opposite charge system. Note, however, that (\ref{A4}), (\ref{A8}) refer to the
combination of the short range (nuclear) and Coulomb (external) interactions,
and the latter acts both in the initial and final states $(\varphi^-_\beta $
and $\varphi^+_\alpha$ in (\ref{A2})). In the more general case, when initial
and final states are different (i.e. short range reaction with different
particles and interactions in the initial and final states), one should keep
track of indices $\beta$ and $\alpha$, which are different, and
$\varphi^-_\beta$ and $\varphi^+_\alpha$ satisfy equations of the form
(\ref{A3}) for $\varphi^-_\beta$ with $U\equiv U_\beta , $ $ K=K_\beta$ and
$\varphi^+_\alpha$ satisfies\be \varphi^+_\alpha = \chi_\alpha + \frac{1}{E-
K_\alpha+ i\varepsilon} U_\alpha \varphi_\alpha^+.\label{A9}\ee

To clarify this situation we consider, as in \cite{19}, a simple
model for the internal interaction \be V(\ver,\ver') =\lambda
g_\beta (r) g_\alpha (r')\label{A10}\ee and as a result the
reaction amplitude becomes \be f_{\beta\alpha} = \lambda I_\beta
I_\alpha, ~~ I_\beta = \int \varphi^-*_\beta g_\beta
d^3r.\label{A11}\ee

Taking into account different radii of internal and external
motion, one can approximate \be I_\beta = \varphi^-*_\beta (0)
\int g_\beta d^3 r, ~~ I_\alpha = \varphi^+_\alpha (0) \int
g_\alpha d^3 r\label{A12}\ee a generalization of
$f_{\alpha\alpha}$ to the case of $\alpha\neq \beta$ is (to the
first order in $f_{in}$) \be f_{\beta\alpha} = \left( \eta^{(\beta
*)}_{Jost}\right)^{-1} f_{in} (E)\left(
\eta^{(\alpha)}_{Jost}\right)^{-1}.\label{A13}\ee we are
especially interested in the case, when both Coulomb interaction
and external magnetic field are present simultaneously. The first
interaction is effective when $\gamma = \left|
\frac{Z\alpha}{v}\right|\sim 1$ and for $\gamma\ll 1$ one can
retain only m.f. effects. In this case both $\varphi^-_\beta$ and
$\varphi^+_\alpha$ are product of a plane wave in the $z$
direction and bound state eigenfunction in the $(x,y)$ plane, \be
\varphi^-_\beta = \varphi^+_\alpha = \frac{e^{ik_zz}}{\sqrt{L}}
\varphi_{n_\bot} (\vex_\bot).\label{A14}\ee

\end{document}